\documentclass[prb,aps,amssymb,showpacs,twocolumn]{revtex4}
\usepackage{amsmath}
\usepackage{amssymb}
\usepackage{amsthm}
\usepackage{amsfonts}
\usepackage{enumerate}
\usepackage{latexsym}
\usepackage{graphicx}

\newcommand{\beq}{\begin{equation}}
\newcommand{\eneq}{\end{equation}}

\input{epsf}

\begin{document}

\tolerance 10000

\newcommand{\vk}{{\bf k}}


\title{An Exact SU(2) Symmetry and Persistent Spin Helix in a Spin-Orbit Coupled System}
\author{B. Andrei Bernevig$^{1,2}$, J. Orenstein$^{3,4}$ and Shou-Cheng Zhang$^{1}$}

\affiliation{$^1$ Department of Physics, McCullough Building,
Stanford University, Stanford, CA 94305-4045} \affiliation{$^2$
Kavli Institute for Theoretical Physics, University of California,
Santa Barbara, CA 93106} \affiliation{$^3$ Lawrence Berkeley
National Laboratory, Berkeley, CA 94720 and} \affiliation{$^4$
Physics Department, University of California at Berkeley, Berkeley,
CA 94720}

\begin{abstract}
Spin-orbit coupled systems generally break the spin rotation
symmetry. However, for a model with equal Rashba and Dresselhauss
coupling constant (the ReD model), and for the $[110]$ Dresselhauss
model, a new type of $SU(2)$ spin rotation symmetry is discovered.
This symmetry is robust against spin-independent disorder and
interactions, and is generated by operators whose wavevector depends
on the coupling strength. It renders the spin lifetime infinite at
this wavevector, giving rise to a Persistent Spin Helix (PSH). We
obtain the spin fluctuation dynamics at, and away, from the symmetry
point, and suggest experiments to observe the PSH.
\end{abstract}

\date{\today}


\maketitle

The physics of systems with spin-orbit coupling has generated great
interest from both academic and practical perspectives
\cite{wolf2001}. In particular, spin-orbit coupling allows for
purely electric manipulation of the electron spin
\cite{nitta1997,grundler2000,kato2004A,kato2004}, and could be of
use in areas from spintronics to quantum computing. Theoretically,
spin-orbit coupling reveals fundamental physics related to
topological phases and their applications to the intrinsic and
quantum spin Hall
effect\cite{murakami2003,sinova2004,kane2005,bernevig2006,qi2005}.

While strong spin-orbit interaction is useful for manipulating the
electron spin by electric fields, it is also known to have the
undesired effect of causing spin-decoherence
\cite{dyakonov1986,pikustitkov}. The decay of spin polarization
reflects the nonconservation of the total spin operator, $\vec{S}$,
i.e. $[\vec{S}, {\cal{H}}] \ne 0$, where ${\cal{H}}$ is any
Hamiltonian that contains spin-orbit coupling. In this Letter, we
identify an exact $SU(2)$ symmetry in a class of spin-orbit systems
that renders the spin lifetime infinite. The symmetry involves
components of spin at a finite wave vector, and is different from
the $U(1)$ symmetry that has previously been associated with this
class of models \cite{schliemann2003,hall2003}. As a result of this
symmetry, spin polarization excited at a certain ``magic" wavevector
will persist. If this symmetry can be realized experimentally, it
may be possible to manipulate spins through spin-orbit coupling
without spin-polarization decay, at length scales characteristic of
today's semiconductor lithography.

We consider a two-dimensional electron gas without inversion
symmetry,  allowing spin-orbit coupling that is linear in the
electron wavevector.  The most general form of linear coupling
includes both Rashba and Dresselhaus contributions:
\begin{equation}
{\cal{H}}= \frac{k^2}{2 m} + \alpha (k_y \sigma_x - k_x \sigma_y) +
\beta (k_x \sigma_x - k_y \sigma_y), \label{rashbaanddresshamilt}
\end{equation}
\noindent where $k_{x,y}$ is the electron momentum along the $[100]$
and $[010]$ directions respectively, $\alpha$, and $\beta$ are the
strengths of the Rashba, and Dresselhauss spin-orbit couplings and
$m$ is the effective electron mass. We shall be interested in the
special case of $\alpha =\beta$, which may be experimentally
accessible through tuning of the Rashba coupling via externally
applied electric fields \cite{nitta1997}. When $\alpha=\beta$, the
spin-orbit coupling part of the Hamiltonian reads $\alpha ( k_x
+k_y) (\sigma_x - \sigma_y)$. Rotating the spatial coordinates by
introducing $k_\pm=\frac{1}{\sqrt{2}}(k_y\pm k_x)$, and also
performing the global spin rotation generated by $U =
\frac{1}{\sqrt{2}}(1 + \frac{i}{\sqrt{2}}(\sigma_x + \sigma_y))$,
brings the Hamiltonian to the diagonal form:
\begin{equation}
{\cal{H_{\rm ReD}} }=  U^\dagger {\cal{H}} U =
\frac{k_+^2+k_-^2}{2m} - 2 \alpha k_{+} \sigma_z.
\label{rashbaequaldresshamilt}
\end{equation}
\noindent  Throughout this paper, we shall be using both the
original spin basis and the transformed spin basis for the
$\cal{H_{\rm ReD}}$ model, depending on the context. It should
always be remembered that $\sigma_z$ in the transformed spin basis
corresponds to $\frac{1}{\sqrt{2}}(\sigma_x - \sigma_y)$ in the
original spin basis. $\cal{H_{\rm ReD}}$ is mathematically
equivalent to the Dresselhauss $[110]$ model, describing quantum
wells grown along the $[110]$ direction, whose Hamiltonian is given
by:
\begin{equation}
{\cal{H_{\rm[110]}} }= \frac{k_x^2+k_y^2}{2m} - 2 \alpha k_x
\sigma_z. \label{[110]}
\end{equation}

\begin{figure}[h]
\includegraphics[scale=0.44]{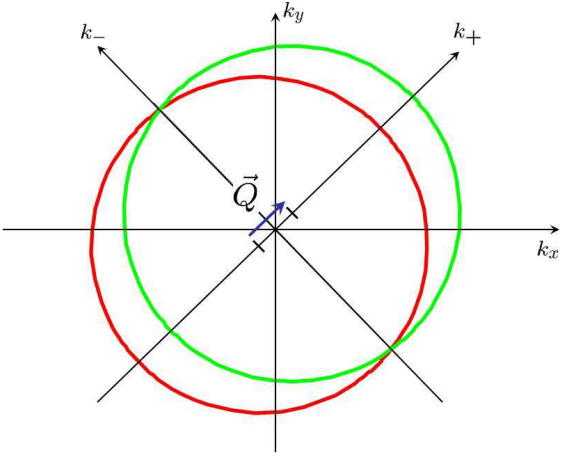}
\caption{[Color Online]  Fermi surfaces of the model consists of two
circles shifted by the wavevector $\vec{Q}= (Q_+, Q_-) =( 4 m
\alpha,0)$. } \label{nesting}
\end{figure}

As the Hamiltonians ${\cal{H_{\rm[ReD]}}}$ and
${\cal{H_{\rm[110]}}}$ are already diagonal, the energy bands (for
${\cal{H_{\rm[ReD]}}}$) are simply given by:
\begin{equation}
\epsilon_{ \downarrow,\uparrow}(\vec{k})= \frac{k^2}{2m} \pm 2
\alpha k_{+}, \label{bandswithnesting}
\end{equation}
\noindent where $\downarrow,\uparrow$ are the spin components in the
new, unitary transformed, spin basis: $\vec{S} \rightarrow U^\dagger
\vec{S} U$. The bands in Eq.[\ref{bandswithnesting}] have an
important {\it shifting property}:
\begin{equation}
\epsilon_{\downarrow}(\vec{k}) = \epsilon_{\uparrow}(\vec{k}
+\vec{Q}),\label{shifting}
\end{equation}
\noindent where $Q_+=4 m \alpha, Q_-=0$ for the
${\cal{H_{\rm[ReD]}}}$ model and $Q_x=4 m \alpha, Q_y=0$ for the
${\cal{H_{\rm[110]}}}$ model. The Fermi surfaces consist of two
circles shifted by the ``magic" shifting vector $\vec{Q}$, as shown
in Fig.[\ref{nesting}].

${\cal{H_{\rm[ReD]}}}$ and ${\cal{H_{\rm[110]}}}$ have the
previously known $U(1)$ symmetry generated by $\sigma_z$
\cite{schliemann2003,hall2003}, which ensures long lifetime for
uniform spin polarization along the z-axis\cite{ohno1999}. The exact
$SU(2)$ symmetry discovered in this work is generated by
 the  following operators, expressed in the transformed spin
basis as:
\begin{eqnarray}
& S^-_{Q} = \sum_{\vec{k}} c^\dagger_{\vec{k} \downarrow} c_{\vec{k}
+ \vec{Q} \uparrow}, \;\;\; S^+_{Q} = \sum_{\vec{k}}
c^\dagger_{\vec{k}+\vec{Q}, \uparrow} c_{\vec{k} \downarrow}
\nonumber \\ & S^z_0 = \sum_{\vec{k}} c^{\dagger}_{\vec{k} \uparrow}
c_{\vec{k} \uparrow} - c^\dagger_{\vec{k} \downarrow} c_{\vec{k}
\downarrow}, \label{su2symmetry}
\end{eqnarray}
\noindent with $c_{k \uparrow, \downarrow}$  being the annihilation
operators of spin-up and down particles. These operators obey the
commutation relations for angular momentum,
\begin{eqnarray}
[S_0^z,S^\pm_Q] = \pm 2 S_Q^\pm; \ \ [S^+_Q, S^-_Q] = S_0^z
\end{eqnarray}
\noindent

The shifting property Eq.[\ref{shifting}] ensures that the operators
defined in Eq.[\ref{su2symmetry}] commute with the Hamiltonian,
\begin{equation}
[{\cal{H_{\rm ReD}}}, c^\dagger_{\vec{k} +\vec{Q} \uparrow}
c_{\vec{k} \downarrow} ]  = (\epsilon_{\uparrow}(\vec{k} + \vec{Q})-
\epsilon_\downarrow (\vec{k})) c^\dagger_{\vec{k} +\vec{Q} \uparrow}
c_{\vec{k} \downarrow} = 0
\end{equation}
\noindent and similarly for $ c^\dagger_{\vec{k} \downarrow}
c_{\vec{k} + \vec{Q} \uparrow}$, thus uncovering the $SU(2)$
symmetry. This symmetry is robust against both spin-independent
disorder and Coulomb (or other many-body) interactions as the spin
operators commute with the finite-wave vector particle density
$\rho_q = \sum_k c^\dagger_{\vec{k} + \vec{q}} c_{\vec{k}}$:
\begin{equation}
[\rho_q, S_Q^{\pm}] = [\rho_q, S^z_0] = 0.
\end{equation}
\noindent As a result single-particle potential scattering terms
like $\sum_q V_q \rho_q$, as well as many-body interaction terms
like $\sum_q V_q \rho_q^\dagger \rho_q$, all commute with the
generators (\ref{su2symmetry}), and the $SU(2)$ symmetry is robust
against these interactions.

\begin{figure}[h]
\includegraphics[scale=0.40]{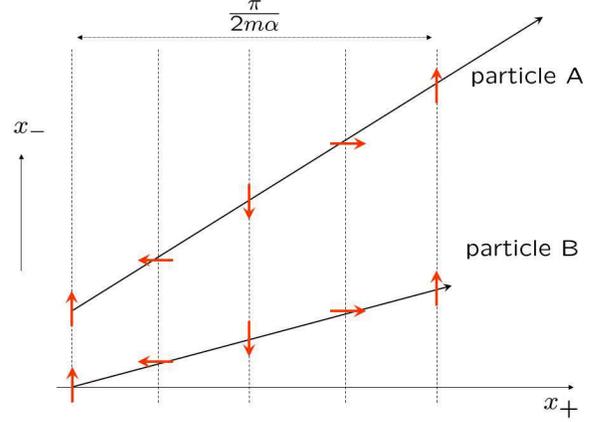}
\caption{ [Color Online] Spin configurations for two particles A and
B depend only on the distance traveled along the $x_+$ axis and not
on their initial momenta. For the same $x_+$ distance traveled, the
spin precesses by exactly the same angle. After a length $L_+ = 2
\pi \hbar/Q_+$ the spins all return exactly to the original
configuration.} \label{precessionangle}
\end{figure}

The $SU(2)$ conservation laws imply that the expectation values of
$S^z$, $S_Q$, and $S_Q^\dagger$ have infinite lifetime. Since $S^z$
is a conserved quantity,  a fluctuation of the $z$-component of spin
polarization with wavevector $q$ can only decay by diffusion, which
takes time $\tau_s=1/D_s q^2$, where $D_s$ is the spin diffusion
constant.  Notice that the hermitian operators $S_x(Q)=
\frac{1}{2}(S_Q^\dag + S_Q)$ and $S_y(Q)= \frac{1}{2i}(S_Q^\dag -
S_Q)$ create a helical spin density wave in which the direction of
the spin polarization rotates in the $x,y$ plane in the transformed
spin basis. Another manifestation of this symmetry is the vanishing
of spin-dependent quantum interference (or weak anti-localization)
at the $SU(2)$ point \cite{pikus1995}.

The infinite lifetime of the $S_x, S_y$ spin helix with wave vector
$\vec Q$ results from the combined effects of diffusion and
precession in the spin-orbit effective field.  A physical picture
for the origin of the persistent spin helix (PSH) is sketched in
Fig.[\ref{precessionangle}],  expressed in the transformed spin
basis. Consider a particle propagating in the plane with momentum
$\vec k$. In time $t$, it travels a distance along the $x_+$
direction $L_+=k_{+}t/m$, while its spin precesses about the $z$
axis by an angle $\theta=4\alpha k_+ t/\hbar$. Eliminating $t$ from
the latter equation yields $\theta= 4\alpha m L_+ /\hbar$,
demonstrating that the net spin precession in the $x,y$ plane
depends only on the net displacement  in the $x_+$ direction and is
independent of any other property of the electron's path. Electrons
starting with parallel spin orientation and the  same value of $x_+$
will return exactly to the original orientation after propagating
$L_+=2\pi\hbar/Q_+$.

\begin{figure}[h]
\includegraphics[scale=0.44]{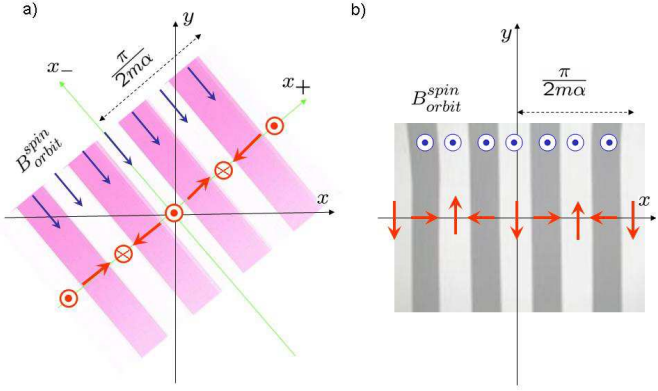}
\caption{ [Color Online] (a) PSH for the ${\cal{H_{\rm ReD}}}$
model. The spin-orbit magnetic field is in-plane (blue), whereas the
spin helix, for the choice of the relative signs in
Eq.[\ref{rashbaanddresshamilt}] is in the $(x_+, z)$ plane. (b) PSH
for the ${\cal{H_{\rm [110]}}}$ model. The spin-orbit magnetic field
$B^{spin}_{orbit}$, in blue, is out of plane, whereas the spin
helix, in red, is in-plane. } \label{spingrating}
\end{figure}

For sake of clarity, we have depicted the PSH in
Fig.[\ref{spingrating}] in the original basis. For a range of values
in different materials from weak to strong spin-orbit splitting
$\alpha = 10^2 \rightarrow 10^4 (m/s)$, the characteristic
wavelength of the PSH is $10 \mu m \rightarrow 100 n m$. In GaAs,
the typical value is $L_+ = 1 \mu m$. This characteristic wave
length is on the scale of typical lithographic dimensions used in
today's semiconductor industry.

Mathematically, the PSH is a direct manifestation of a
``non-abelian" flux in the ground state of the ${\cal{H_{\rm ReD}}}$
and the ${\cal{H_{\rm [110]}}}$ models. Following Ref.
(\onlinecite{jin2006}), we express ${\cal{H_{\rm ReD}}}$ in the form
of a background non-abelian gauge potential,
\begin{equation}
{\cal{H_{\rm ReD}}} = \frac{k_-^2}{2m} + \frac{1}{2m}(k_+ - 2 m
\alpha \sigma_z)^2 + const.
\end{equation}
\noindent In contrast to the general case where the non-abelian
gauge potential leads to a finite non-abelian field strength, the
field strength vanishes identically for $\alpha=\beta$. Therefore,
we can eliminate the vector potential by a non-abelian gauge
transformation: $\Psi_{\uparrow}(x_+,x_-) \rightarrow \exp( i 2 m
\alpha x_+) \Psi_{\uparrow}(x_+,x_-)$, $\Psi_{\downarrow}(x_+,x_-)
\rightarrow \exp(-  i 2 m \alpha x_+) \Psi_{\downarrow}(x_+,x_-)$.
Under this transformation, the spin-orbit coupled Hamiltonian is
mapped to that of the free Fermi gas. The cost of the transformation
is two-fold. First, the new wave function $\Psi (x_+,x_-)$ satisfies
twisted spin boundary conditions\cite{qi2006}. However, these have
no physical effect, since a metallic system without
off-diagonal-long-range-order is insensitive to a change of boundary
conditions in the thermodynamic limit. Second, while diagonal
operators such as the charge $n$ and $S_z$  remain unchanged,
off-diagonal operators, such as $S^-(\vec{x}) =
\psi_\downarrow^\dagger(\vec{x}) \psi_\uparrow (\vec{x})$ and
$S^+(\vec{x}) = \psi_\uparrow^\dagger(\vec{x}) \psi_\downarrow
(\vec{x})$ are transformed: $S^-(\vec{x}) \rightarrow \exp(- i 4 m
\alpha x_+) S^-(\vec{x})$, $S^+(\vec{x}) \rightarrow \exp( i 4 m
\alpha x_+) S^+(\vec{x})$. In the transformed basis, all three
components of the spin obey the simple diffusion equation, as the
Hamiltonian is just that for free electrons without spin-orbit
coupling. Hence $S_x = const$ is a solution to the diffusion
equation. Transforming back demonstrates that this solution
corresponds to
 $(S_x, S_y) =(\cos(4m \alpha x_+), \sin(4 m \alpha
x_+))=const$ ., which is the PSH.

While spin is conserved at the $SU(2)$ point, dynamics emerge when
the conditions $\vec{q}=\vec{Q}$ and/or $\alpha=\beta$ are not met.
Solving for the coupled charge and spin dynamics requires Boltzman
transport equations, which we obtain below using the Keyldish
formalism \cite{rammer1986,mishchenko2004}. Assuming isotropic
scattering with momentum lifetime $\tau$, the retarded and advanced
Green's functions are:
\begin{equation}
G^{R, A}(k, \epsilon) = (\epsilon - {\cal{H}} \pm
\frac{i}{2\tau})^{-1}.
\end{equation}
\noindent We introduce a momentum, energy, and position dependent
charge-spin density which is a $2\times 2$ matrix $g(k, r,t)$.
Summing over momentum:
\begin{equation}
\rho({ r, t}) \equiv\int \frac{d^2 k}{(2 \pi)^{3} \nu} g({k,r,t}),
\end{equation}
\noindent gives the real-space spin-charge density  $\rho(r,t) =
n(r,t) + S^i(r,t) \sigma_i $, where $n(r,t)$ and $S^i(r,t)$ are the
charge and spin density and $\nu ={m}/{2 \pi} $ is the density of
states in two-dimensions. $\rho(r,t) $ and $g(k, r,t)$ satisfy a
Boltzman-type equation \cite{rammer1986,mishchenko2004}:
\begin{equation}
\frac{\partial g}{\partial t} + \frac{1}{2} \left\{\frac{\partial
{\cal{H}}}{\partial k_i}, \frac{\partial g}{\partial r_i} \right\} +
i \left[{\cal{H}}, g \right] = - \frac{g}{\tau} + \frac{i}{\tau}(G^R
\rho - \rho G^A).
\end{equation}
\noindent that we now solve for the Hamiltonian of
Eq.[\ref{rashbaanddresshamilt}] for arbitrary $\alpha, \beta$. To
obtain the spin-charge transport equations, we follow the sequence:
time-Fourier transform the above equation; find a general solution
for $g(k,r,t)$ involving $\rho(r,t)$ and the $k$-dependent
spin-orbit coupling; perform a gradient expansion of that solution
(assuming $\partial_{r} << k_F$ where $k_F$ is the Fermi wavevector)
to
 second order;
and, finally, integrate over the momentum. The anisotropic nature of
the Fermi surfaces when both $\alpha$ and $\beta$ are nonzero
introduces  coupling of charge and spin degrees of freedom beyond
what was found with only Rashba coupling \cite{burkov2003}.  The
final result, expressed in the $i = (x_+, x_-, z)$ spatial
coordinates and the original spin basis, is:
\begin{equation}
\partial_t n  =  D \partial_{i}^2 n +  B_1 \partial_{x_+} S_{x_-} -  B_2 \partial_{x_-}
S_{x_+} \label{syequation}
\end{equation}
\begin{equation}
\partial_t S_{x_-}  = D \partial_{i}^2  S_{x_-} +  B_{1}  \partial_{x_+} n
- C_2 \partial_{x_-} S_z  - T_2 S_{x_- } \label{sxequation}
\end{equation}
\begin{equation}
\partial_t S_{x_+}  =  D \partial_{i}^2   S_{x_+} -   B_2 \partial_{x_-} n
- C_{1} \partial_{x_+} S_z  - T_1  S_{x_+} \label{syequation}
\end{equation}
\begin{equation}
\partial_t S_z =  D \partial_i^2 S_z +   C_2 \partial_{x_-} S_{x_-} +  C_1
\partial_{x_+} S_{x_+} - (T_1 +T_2) S_z. \label{szequation}
\end{equation}
\noindent With an effective  $k_F$ defined as  $\sqrt{2m E_F}$, the
constants in the above equations are:
\begin{eqnarray}
& B_1= 2 (\alpha - \beta)^2(\alpha + \beta) k_F^2 \tau^2, \nonumber
\\ &
B_2= 2 (\alpha + \beta)^2(\alpha - \beta) k_F^2 \tau^2, \nonumber \\
& C_1= 2\frac{1}{m} (\alpha + \beta) k_F^2 \tau, \;\;\;\; C_2=
2\frac{1}{m} (\alpha - \beta) k_F^2 \tau \nonumber \\ & T_1 =
2(\alpha+ \beta)^2 k_F^2 \tau, \;\;\;\; T_2 = 2(\alpha- \beta)^2
k_F^2 \tau.
\end{eqnarray}
\noindent The diffusion constant $D = v_F^2 \tau/2$. Our results for
the coupling coefficients  are valid in the diffusive limit $\alpha
k_F \tau  << 1,\; \beta k_F \tau  << 1$ and reduce to the
appropriate limits in the cases $\beta \rightarrow 0$ or $\alpha
\rightarrow 0$ \cite{burkov2003,mishchenko2004}.  We observe that
for a general direction of propagation in the $x_\pm$ plane, the
three components of spin, and the charge, are all coupled.

Motivated by experiments that probe spin dynamics optically, we
focus on the behavior of the out-of-plane component of the spin,
which is $S_z$ in the original basis. We assume a space-time
dependence proportional to $\exp[i(\omega t -
\vec{q}\cdot\vec{r})]$, and compare $\vec{q}$ parallel to
$[1\bar{1}0]$ and $\vec{q}$ parallel to
 $[110]$.

The spin polarization lifetime for $\vec{q}$ parallel to
$[1\bar{1}0]$ is not enhanced, as no shifting property exists in
this direction. For this orientation, the four equations [14-17]
separate into two coupled pairs, with $n$ being coupled  to
$S_{x_+}$ and  $S_z$ coupled  to $S_{x_-}$. The optical probe will
detect the characteristic frequencies that are solutions to the
equation that couples $S_z$ with $S_{x_-}$:

\begin{equation}
i \omega_{1,2} =  - D q^2 - \frac{1}{2} (2 T_2 + T_1 \pm \sqrt{T_1^2
+ 4 q^2 C_2^2}).
\end{equation}
\noindent At $\alpha = \beta$ we have $T_2 = C_2=0$ and the
characteristic frequencies become $i \omega_{1} = - D q^2- T_1$, $i
\omega_{2} = - D q^2$.

The spin polarization dynamics are qualitatively different in the
$[110]$ direction, which is the direction along which the Fermi
surfaces are shifted. In this
 case the four equations
again decouple into two pairs, with $n$ coupled to $S_{x_-}$ and
$S_z$ coupled  to $S_{x_+}$. The characteristic frequencies are:
\begin{equation}
i \omega_{1,2} =  - D q^2 - \frac{1}{2} (2 T_1 + T_2 \pm \sqrt{T_2^2
+ 4 q^2 C_1^2}).
\end{equation}
\noindent We note that $i\omega_2(q)$ has a minimum at a wavevector
that depends only on $\alpha, \beta$, and $m$, $i.e.$, is
independent of $k_F$ and $\tau$.  When $\alpha =\beta$ the decay
rates become $i \omega_{1,2}= - D q^2- T_1 \mp C_1 q$, and
$i\omega_2(q)$ is zero (corresponding to the PSH) at the shifting
wave-vector $4m \alpha$. We have checked that this relation
continues to hold as we include higher order corrections to the
transport coefficients. Earlier calculations based on the pure
Rashba model \cite{froltsov2001,burkov2003} also predict an
increased spin lifetime at a finite wave vector $(\sqrt{15}/{2})
\alpha m$, but the lifetime is enhanced relative to $q=0$ only by
the factor $16/7$.

Transient spin-grating experiments \cite{cameron1996,weber2005} are
particularly well-suited to testing our theoretical predictions
(Eqs.[19,20]) and discovering the PSH, as they inject finite
wavevector spin distributions. If at $t=0$ an $S_z$ polarization
proportional to $\cos(q x_+)$ is excited, we predict a time
evolution,
\begin{equation}
S_z(q,t) = A_1(q) e^{i \omega_1(q)t} + A_2(q) e^{i \omega_2(q) t},
\end{equation}
 \noindent where
\begin{equation}
A_{1,2}(q) = \frac{1}{2} \left[ 1 \pm \frac{T_2}{\sqrt{T_2^2 + 4 q^2
C_1^2}}\right].
\end{equation}
\noindent Thus, according to theory, the photoinjected spin
polarization wave will decay as a double exponential, with
characteristic decay rates $i\omega_{1,2}(q)$. As $\alpha
\rightarrow \beta$ the weight factors of the two exponentials become
equal and the decay rate of the slower component $\rightarrow 0$ at
the ``magic" wave-vector. If transient grating measurements verify
these predictions, they will be able to provide rapid and accurate
determination of the spin-orbit Hamiltonian, enabling tuning of
sample parameters to achieve $\alpha=\beta$.

In conclusion, we have discovered a new type of spin $SU(2)$
symmetry in a class of spin-orbit coupled models including the ReD
model and the Dresselhauss $[110]$ model. Based on this symmetry, we
predict the existence of a Persistent Spin Helix. The lifetime of
the PSH is infinite within these models, but of course there will be
other relaxation mechanisms which would lead to its eventual decay.
We obtained the transport equations for arbitrary strength of Rashba
and Dresselhauss couplings and these equations provide the basis to
analyze experiments in search of the PSH.

B.A.B. wishes to acknowledge the hospitality of the Kavli Institute
for Theoretical Physics at University of California at Santa
Barbara, where part of this work was performed. This work is
supported by the NSF through the grants DMR-0342832, by the US
Department of Energy, Office of Basic Energy Sciences under contract
DE-AC03-76SF00515, the Western Institute of Nano-electronics and the
IBM Stanford SpinApps Center.

\end{document}